\documentstyle{article}
\begin{document}

\baselineskip=23pt

\begin{flushleft}
{\bf {\Huge
Correlations between various hardness ratios of gamma-ray bursts
}}\\
\vspace{4mm} 
{\bf Yi-Ping Qin$^{1,2,3,4}$, Guang-Zhong Xie$^{1,2,4}$, Xue-Tang Zheng$^{5}$,
and En-Wei Liang$^{1,2,6}$,}

{\bf $^{1}$ Yunnan Observatory, Chinese Academy of Sciences, Kunming, Yunnan
650011, P. R. China}

{\bf $^{2}$ National Astronomical Observatories, Chinese Academy of Sciences 
}

{\bf $^{3}$ Chinese Academy of Science-Peking University joint Beijing
Astrophysical Center }

{\bf $^{4}$ Yunnan Astrophysics Center, Yunnan University, Kunming, Yunnan
650091, P. R. China}

{\bf $^{5}$Department of Physics, Nanjing University of Science and
Technology, Nanjing, Jiangsu 210014, P. R. China}\\

{\bf $^{6}$Physics Department, Guangxi University, Nanning, Guangxi 530004,
P. R. China}\\

\vspace{10mm} 

{\bf {\Large ABSTRACT}}

\end{flushleft}

We study correlations between various hardness ratios of gamma-ray bursts
(GRBs) and investigate if there are any differences between the two classes
of the objects in the distributions of the ratios. The study shows that: (1)
almost all hardness ratios of GRBs are mutually correlated; (2) the most
obvious correlation for both the short and long duration classes is that
between $\log HR_{41}$ and $\log HR_{42}$, where $HR_{ij}$ is the fluence in
channel $i$ divided by the fluence in channel $j$; (3) for the short
duration class, $\log HR_{21}$ and $\log HR_{32}$ as well as $\log HR_{31}$
and $\log HR_{43}$ are not correlated, while for the long duration class, $%
\log HR_{21}$ and $\log HR_{42}$ as well as $\log HR_{21}$ and $\log HR_{43}$
are not correlated; (4) those most significant correlations for the long
duration class are more obvious than that for the short duration class; (5)
the data of hardness ratios for the long duration class are more
concentrated, while that for the short duration class are more scattered;
(6) the sources of the two classes are distributed in different areas in the
hardness ratio---hardness ratio plot. These results suggest that,
statistically, the slope of the higher part of the spectrum of the long
duration bursts has nothing to do with that of the lower part; emissions at
higher energy bands from the bursts of both short and long duration classes
must be significantly different for different sources, while radiations at
lower energy bands of the objects are relatively similar; the spectrum of
the short duration bursts must be harder than that of the long duration
bursts, confirming what the well-known hardness-duration correlation
reveals; the profiles of the spectra between the long duration bursts must
be more similar than that between the short duration bursts. The long
duration bursts would share more common properties than the short duration
bursts. A possible interpretation is proposed with the concept of the
Doppler boosting in the relativistic beaming model in AGNs.

\begin{flushleft}

{\bf Key words:}
cosmology: observations --- gamma rays: bursts --- gamma rays: theory ---
methods: statistical
\end{flushleft}

\section{INTRODUCTION}

Gamma-ray bursts (GRBs) are a transient astrophysical phenomenon in which
the emission is confined exclusively to high energies: they are detected at
gamma-ray bands and last shortly (from a few milliseconds to several hundred
seconds). Since the discovery of the events about thirty years ago
(Klebesadel et al. 1973), more and more data have been obtained (e.g.,
Fishman et al. 1994; Meegan et al. 1994, 1996, 1998; Paciesas et al. 1997,
1999). Based on the data available so far, statistical studies have recently
unveiled many basic properties of the objects. For example, two separate
classes of GRBs with a division at the duration of 2 s were detected
(Dezalay et al. 1992; Hurley 1992; Kouveliotou et al. 1993); a correlation
between the peak spectral hardness and the peak intensity was discovered
(Atteia et al. 1991, 1994, 1997; Belli 1992; Mitrofanov et al. 1992, 1996;
Paciesas et al 1992; Nemiroff et al. 1994; Mallozzi et al. 1995; Dezalay et
al. 1997).

More recently, the hardness-duration correlation was confirmed by Fishman
(1999) with a large number of bursts observed with BATSE. Generally, the
hardness ratio is defined as the fluence in channel 3 ($\sim 100$ to $\sim
300$ keV) divided by the fluence in channel 2 ($\sim 50$ to $\sim 100$ keV)
(see, e.g., Fishman et al. 1994). This quantity is more likely to present an
intrinsic property of the objects rather than an observational one. It is
noticed that, besides the two channel fluences, there are two other channel
fluences available in the BATSE burst catalogs, e.g., channels 1 ($\sim 20$
to $\sim 50$ keV) and 4 ($>300$ keV) (see, e.g., Paciesas et al. 1997).
Therefore, one can define more than one hardness ratios. We wonder if there
are any meaningful statistical relations between these different hardness
ratios. We also want to make clear if there are any differences between the
two classes of the objects in the distributions of the ratios.

In Section 2, correlations between various hardness ratios of GRBs will be
studied, while in Section 3, differences between the two classes of the
objects in the distributions of the ratios will be investigated.

\section{CORRELATIONS BETWEEN HARDNESS RATIOS}

In the 4B catalog (Paciesas et al. 1997), there are 1027 sources with all of
their four channel (1-4) fluences available. We then get a rather sizable
sample from the catalog (called Sample 1). A hardness ratio $HR_{ij}$ is
defined as 
\begin{equation}
HR_{ij}\equiv f_i/f_j,
\end{equation}
where $i$ and $j$ denote the channels $i$ and $j$ respectively, and $f_i$
and $f_j$ represent the fluences of the corresponding channels respectively.

The plots of $\log HR_{ij}-\log HR_{i^{\prime }j^{\prime }}$ for Sample 1
are shown in Figures 1.1 to 1.15. The correlation coefficients for every
couple of hardness ratios are presented in Table 1.

\vspace{3mm}

\begin{center}
{\bf Table 1. Correlation coefficients for Sample 1 (N=1027)}

\begin{tabular}{cccccc}
\hline
& $\log HR_{31}$ & $\log HR_{32}$ & $\log HR_{41}$ & $\log HR_{42}$ & $\log
HR_{43}$ \\ \hline
{$\log HR_{21}$} & 0.746 & 0.276 & 0.373 & 0.080 & -0.037 \\ 
{$\log HR_{31}$} &  & 0.846 & 0.677 & 0.488 & 0.186 \\ 
{$\log HR_{32}$} &  &  & 0.678 & 0.640 & 0.298 \\ 
{$\log HR_{41}$} &  &  &  & 0.955 & 0.849 \\ 
{$\log HR_{42}$} &  &  &  &  & 0.924 \\ \hline
\end{tabular}
\end{center}

\vspace{3mm}

From Table 1 and Figures 1.1 to 1.15 we find that almost all hardness ratios
are mutually correlated, with only $\log HR_{21}$ and $\log HR_{43}$ being
uncorrelated. The most obvious correlation is that between $\log HR_{41}$
and $\log HR_{42}$. It suggests that, statistically, the slope of the higher
part of the spectrum of most GRBs has nothing to do with that of the lower
part; emissions at higher energy bands from the objects must be
significantly different for different sources, while radiations at lower
energy bands must be relatively similar.

\section{DIFFERENCES BETWEEN THE TWO CLASSES}

To study the differences between the two classes of GRBs in the
distributions of hardness ratios, we simply concern those sources with their
data of the duration as well as the four channel fluences available. The
bursts are divided into two classes with a division at the duration of 2 s:
those sources with $T_{90}\leq 2$ s belong to the short duration class and
those with $T_{90}>2$ s constitute the long duration class, where $T_{90}$
is the time during which the burst integrated counts increases from $5\%$ to 
$95\%$ of the total counts. We find 249 short duration sources (called
Sample 2) and 706 long duration bursts (called Sample 3) from the 4B catalog.

The plots of $\log HR_{ij}-\log HR_{i^{\prime }j^{\prime }}$ for Samples 2
and 3 are shown in Figures 2.1 to 2.15. The correlation coefficients for
every couple of hardness ratios for the samples are presented in Tables 2
and 3.

\vspace{3mm}

\begin{center}
{\bf Table 2. Correlation coefficients for Sample 2 (N=249)}

\begin{tabular}{cccccc}
\hline
& $\log HR_{31}$ & $\log HR_{32}$ & $\log HR_{41}$ & $\log HR_{42}$ & $\log
HR_{43}$ \\ \hline
{$\log HR_{21}$} & 0.653 & -0.050 & 0.206 & -0.218 & -0.235 \\ 
{$\log HR_{31}$} &  & 0.724 & 0.558 & 0.280 & -0.068 \\ 
{$\log HR_{32}$} &  &  & 0.549 & 0.568 & 0.125 \\ 
{$\log HR_{41}$} &  &  &  & 0.910 & 0.790 \\ 
{$\log HR_{42}$} &  &  &  &  & 0.887 \\ \hline
\end{tabular}
\vspace{3mm}

\vspace{3mm} {\bf Table 3. Correlation coefficients for Sample 3 (N=706)}

\begin{tabular}{cccccc}
\hline
& $\log HR_{31}$ & $\log HR_{32}$ & $\log HR_{41}$ & $\log HR_{42}$ & $\log
HR_{43}$ \\ \hline
{$\log HR_{21}$} & 0.757 & 0.308 & 0.324 & 0.059 & -0.060 \\ 
{$\log HR_{31}$} &  & 0.855 & 0.606 & 0.425 & 0.142 \\ 
{$\log HR_{32}$} &  &  & 0.624 & 0.572 & 0.254 \\ 
{$\log HR_{41}$} &  &  &  & 0.964 & 0.874 \\ 
{$\log HR_{42}$} &  &  &  &  & 0.939 \\ \hline
\end{tabular}
\end{center}

\vspace{3mm}

From Tables 2 and 3 we find that, the two classes have similar correlations
between hardness ratios: many of them are mutually correlated, with only a
few exceptions. For the short duration class, $\log HR_{21}$ and $\log
HR_{32}$ as well as $\log HR_{31}$ and $\log HR_{43}$ are not correlated,
while for the long duration class, $\log HR_{21}$ and $\log HR_{42}$ as well
as $\log HR_{21}$ and $\log HR_{43}$ are not correlated, suggesting that,
statistically, the slope of the higher part of the spectrum of the long
duration bursts has nothing to do with that of the lower part. The most
obvious correlation for both the short and long duration classes is that
between $\log HR_{41}$ and $\log HR_{42}$, showing that, statistically,
emissions at higher energy bands from the bursts of both long and short
duration classes must be significantly different for different sources,
while radiations at lower energy bands must be relatively similar.

For the most obvious correlations shown in the two tables, those for the
long duration class are more significant than that for the short duration
class. One can find from Figures 2.1 to 2.15 that the data of hardness
ratios for the long duration class are more concentrated, while that for the
short duration class are more scattered. This may be the most significant
difference between the two classes shown in the figures. It implies
statistically that the profiles of the spectra between the long duration
bursts must be more similar than that between the short duration bursts. The
long duration bursts must share more common properties than the short
duration bursts.

Figures 2.1 to 2.15 also show different distributions of the two classes.
The short duration bursts tend to occupy the right top part of the figures,
showing that, in general, their spectra must be harder than that of the long
duration bursts, confirming what the well-known hardness-duration
correlation reveals (see, e.g., Fishman 1999). This can be verified by a
direct calculation for the two classes. For example, the mean value of $\log
HR_{41}$ for Sample 2 is 1.5 while that for Sample 3 is 0.67.

\section{DISCUSSION AND CONCLUSIONS}

In the last two sections, we study the correlations between various hardness
ratios of GRBs and investigate if there are any differences between the two
classes of the objects in the distributions of the ratios with the data of
the 4B catalog. The results are: (1) almost all hardness ratios of GRBs are
mutually correlated; (2) the most obvious correlation for both the short and
long duration classes is that between $\log HR_{41}$ and $\log HR_{42}$; (3)
for the short duration class, $\log HR_{21}$ and $\log HR_{32}$ as well as $%
\log HR_{31}$ and $\log HR_{43}$ are not correlated, while for the long
duration class, $\log HR_{21}$ and $\log HR_{42}$ as well as $\log HR_{21}$
and $\log HR_{43}$ are not correlated; (4) those most significant
correlations for the long duration class are more obvious than that for the
short duration class; (5) the data of hardness ratios for the long duration
class are more concentrated, while that for the short duration class are
more scattered; (6) the sources of the two classes are distributed in
different areas in the $\log HR_{ij}-\log HR_{i^{\prime }j^{\prime }}$ plot.
These results suggest that, statistically, the slope of the higher part of
the spectrum of the long duration bursts has nothing to do with that of the
lower part; emissions at higher energy bands from the bursts of both short
and long duration classes must be significantly different for different
sources, while radiations at lower energy bands must be relatively similar;
the spectrum of the short duration bursts must be harder than that of the
long duration bursts, confirming what the well-known hardness-duration
correlation reveals; the profiles of the spectra between the long duration
bursts must be more similar than that between the short duration bursts. The
long duration bursts must share more common properties than the short
duration bursts.

According to the most successful model proposed so far, GRBs would result in
relativistically expanding fireballs, and the gamma-ray emission would arise
after the fireball becomes optically thin, in shocks occurring either
because the ejecta run into an external medium or because internal shocks
occur in a relativistic wind (see M\'esz\'aros \& Rees 1997). Several of the
features reported in the first GRB detected over timescales of greater than
days at X-ray and optical wavelengths, i.e., GB 970228 (Costa et al. 1997),
agreed well with the theoretical expectations of the model. In the
well-known relativistic beaming model used in AGNs (Rees 1966, 1967), the
enhancement of the observed flux for a relativistically moving component is $%
\delta ^p$, with $p=3+\alpha $ in the case of a moving sphere and $%
p=2+\alpha $ in the case of a continuous jet, where $\delta $ is the Doppler
factor and $\alpha $ is the spectral index of the component (see, e.g.,
Ghisellini et al. 1993). Recently, the correlation between polarization and
variation for BL Lac objects was found able to be adopted to GRBs and a
value as big as 100 of the Doppler factor for the bursts was estimated by
the relation (see Fan et al. 1997, 1999; Cheng et al. 1999). As mentioned
above, different to that radiated at lower energy bands, emissions of GRBs
at higher energy bands are significantly different for different sources.
This leads to the prospect that the Doppler boosting for GRBs might play a
more important role at higher energy bands than at lower bands. If it is
true, higher energies must be emitted by the fastest moving ejecta while
lower energies be radiated by the slower moving components in both the
isotropic expanding fireball model or the beaming model; or, the range of
the spectral index of GRBs at higher energy bands must be significantly
wider than that at lower energy bands.

\vspace{20mm} 
\begin{flushleft}
{\bf {\Large ACKNOWLEDGMENTS}}\\
\end{flushleft}

This work was supported by the United Laboratory of Optical Astronomy, CAS,
the Natural Science Foundation of China, and the Natural Science Foundation
of Yunnan.

\newpage

\begin{flushleft}
{\bf {\Large REFERENCES}}\\
\end{flushleft}

\begin{verse}
Atteia, J.-L., et al. 1991, Proc. 22d Int. Cosmic-Ray Conf. (Dublin), 93 \\%
Atteia, J.-L., et al. 1994, A\&A, 288, 213 \\Atteia, J.-L., et al. 1997, in
AIP Conf. Proc. 384, ed. C. Kouveliotou, M. F. Briggs, \& G. J. Fishman (New
York: AIP), 301 \\Belli, B. M. 1992, in AIP Conf. Proc. 265, ed. W. S.
Paciesas \& G. J. Fishman (New York: AIP), 100 \\Cheng, K. S., et al. 1999,
astro-ph/9910543 \\Costa, E., et al. 1997, Nature, 387, 783 \\Dezalay,
J.-P., et al. 1992, in AIP Conf. Proc. 265, ed. W. S. Paciesas \& G. J.
Fishman (New York: AIP), 304 \\Dezalay, J.-P., et al. 1997, ApJ, 490, L17 \\%
Fan, J. H., et al. 1997, A\&A 327, 947 \\Fan, J. H., et al. 1999,
astro-ph/9910540 \\Fishman, G. J., 1999, A\&AS, 138, 395 \\Fishman, G. J.,
et al. 1994, ApJS, 92, 229 \\Ghisellini, G., et al. 1993, ApJ, 407, 65 \\%
Hurley, K. C. 1992, in AIP Conf. Proc. 265, ed. W. S. Paciesas \& G. J.
Fishman (New York: AIP), 3 \\Klebesadel, R., Strong, I., Olson, R. 1973,
ApJ, 182, L85 \\Kouveliotou, C., et al. 1993, ApJ, 413, L101 \\Mallozzi, R.
S., et al. 1995, ApJ, 454, 597 \\Meegan, C. A., et al. 1994, The Second
BATSE Burst Catalog, available electronically from the Compton Observatory
Science Support Center \\Meegan, C. A., et al. 1996, ApJS, 106, 65 \\Meegan,
C. A., et al. 1998, in AIP Conf. Proc. 428, Gamma-Ray Bursts: 4th Huntsville
Symp., ed. C. A. Meegan, R. D. Preece, \& T. M. Koshut (New York: AIP), 3 \\%
M\'esz\'aros, P., \& Rees, M. J. 1997, ApJ, 476, 232 \\Mitrofanov, L., et
al. 1992, in Gamma-ray Bursts, ed. C. Ho, R. I. Epstein, \& E. E. Fenimore
(Cambridge: Cambridge Univ. Press), 203 \\Mitrofanov, L., et al. 1996, ApJ,
459, 570 \\Nemiroff, R. J., et al. 1994, ApJ, 435, L133 \\Paciesas, W. S.,
et al. 1992, in AIP Conf. Proc. 265, ed. W. S. Paciesas \& G. J. Fishman
(New York: AIP), 190 \\Paciesas, W. S., et al. 1997, The Fourth BATSE Burst
Catalog, available electronically at
http://cossc.gsfc.nasa.gov/cossc/batse/4Bcatalog \\Paciesas, W. S., et al.
1999, ApJS, 122, 465 \\Rees, M. J. 1966, Nature, 211, 468 \\Rees, M. J.
1967, MNRAS, 135, 345 \\
\end{verse}

\vspace{20mm} 
\begin{flushleft}
{\bf {\Large FIGURE CAPTION}}\\
\end{flushleft}

\begin{verse}
{\bf Figures 1.1 --- 1.15.} The plot of $\log HR_{ij}-\log HR_{i^{\prime
}j^{\prime }}$ for Sample 1.

{\bf Figures 2.1 --- 2.15.} The plot of $\log HR_{ij}-\log HR_{i^{\prime
}j^{\prime }}$ for Samples 2 and 3, where an open circle represents a source
of Sample 2 and a filled circle stands for a source of Sample 3.\\ 
\end{verse}

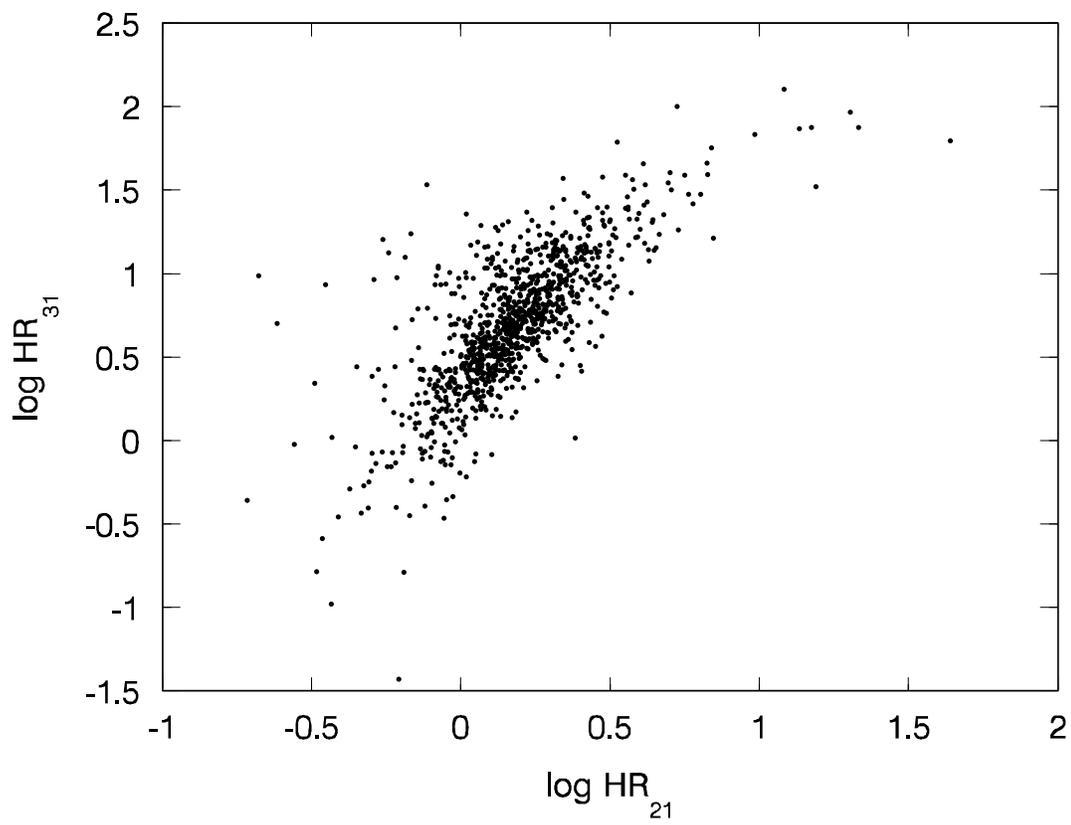
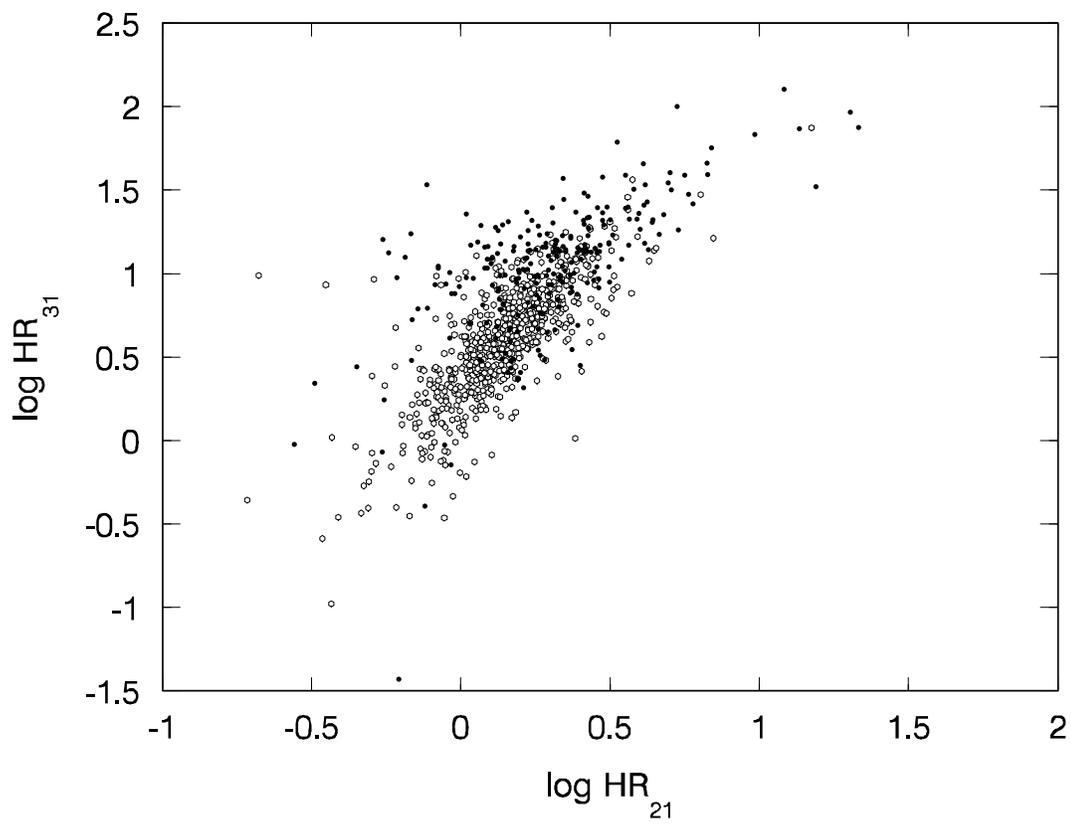
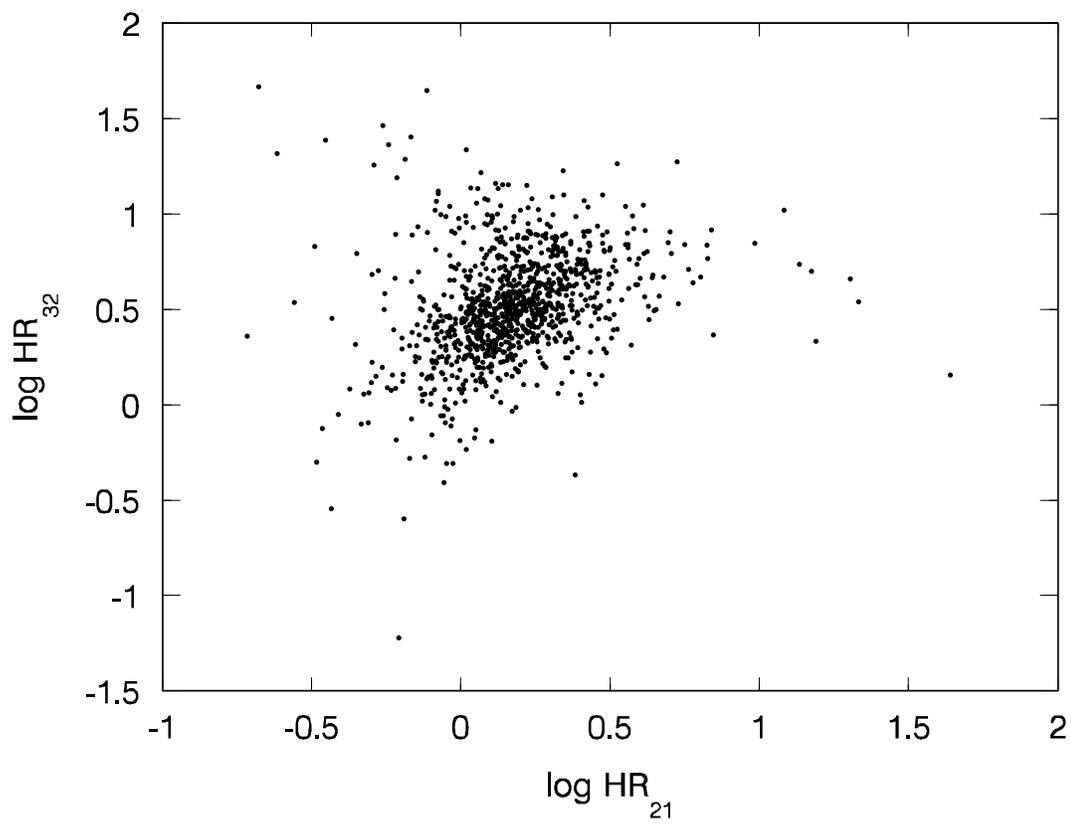
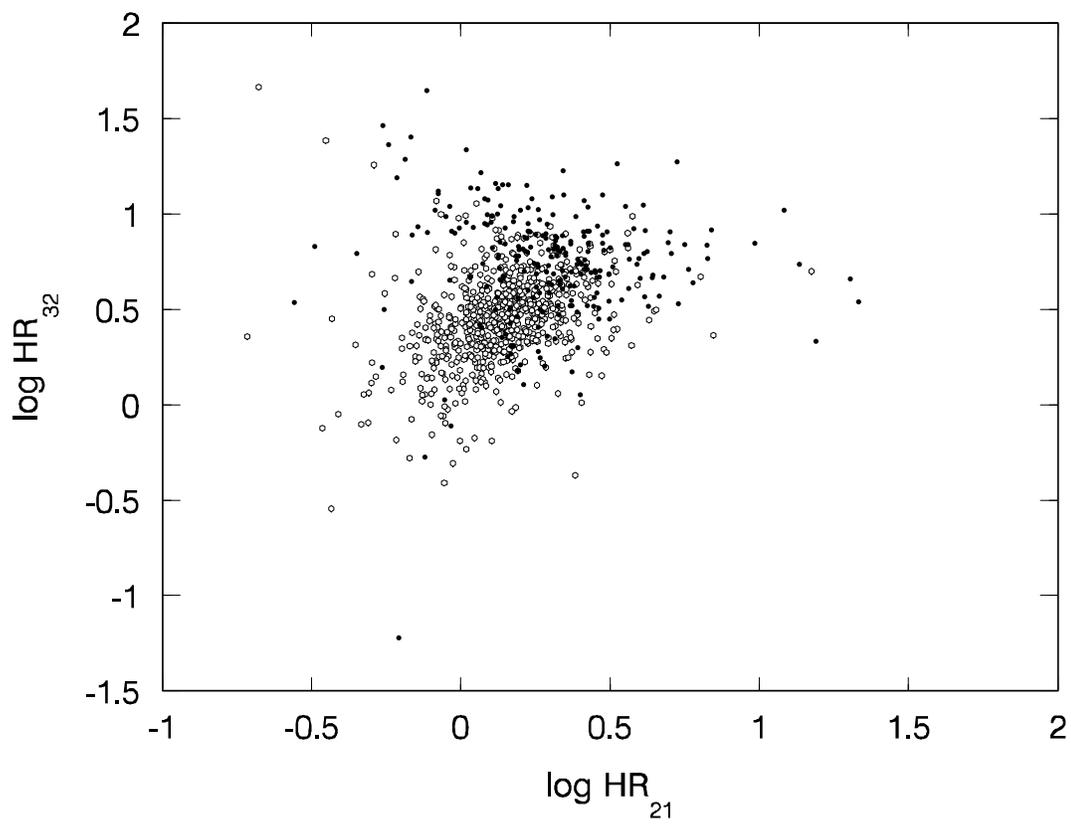
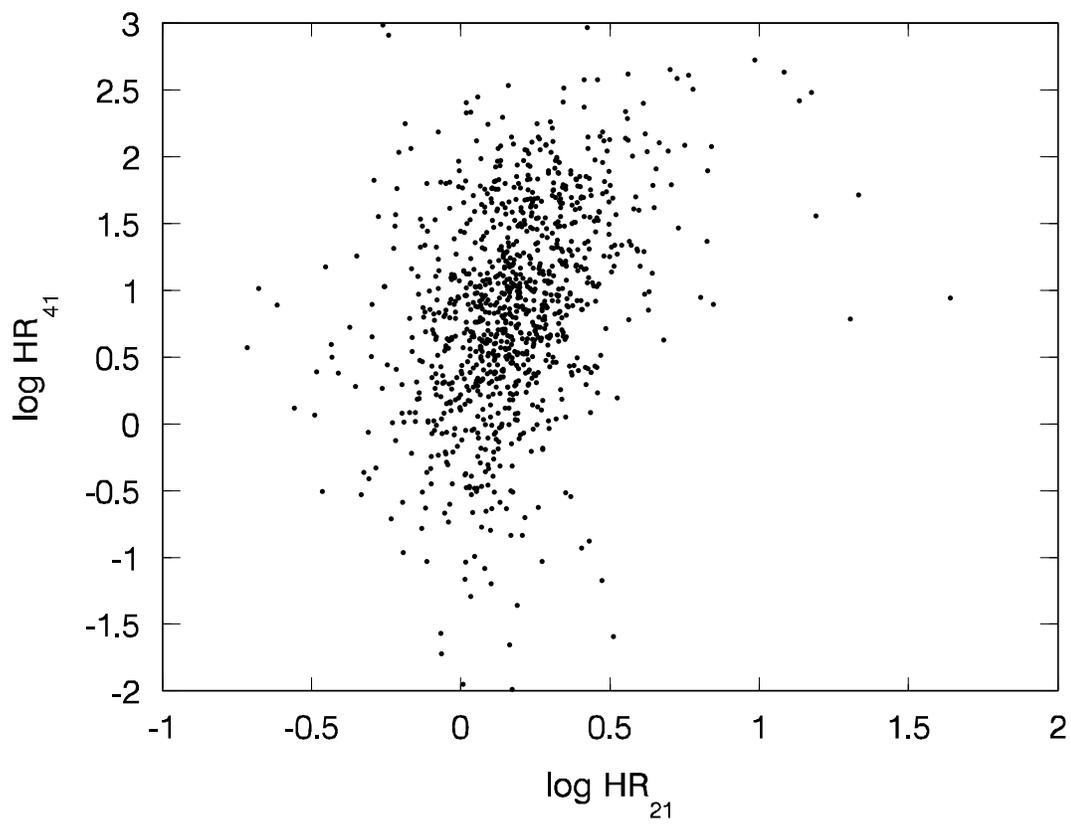
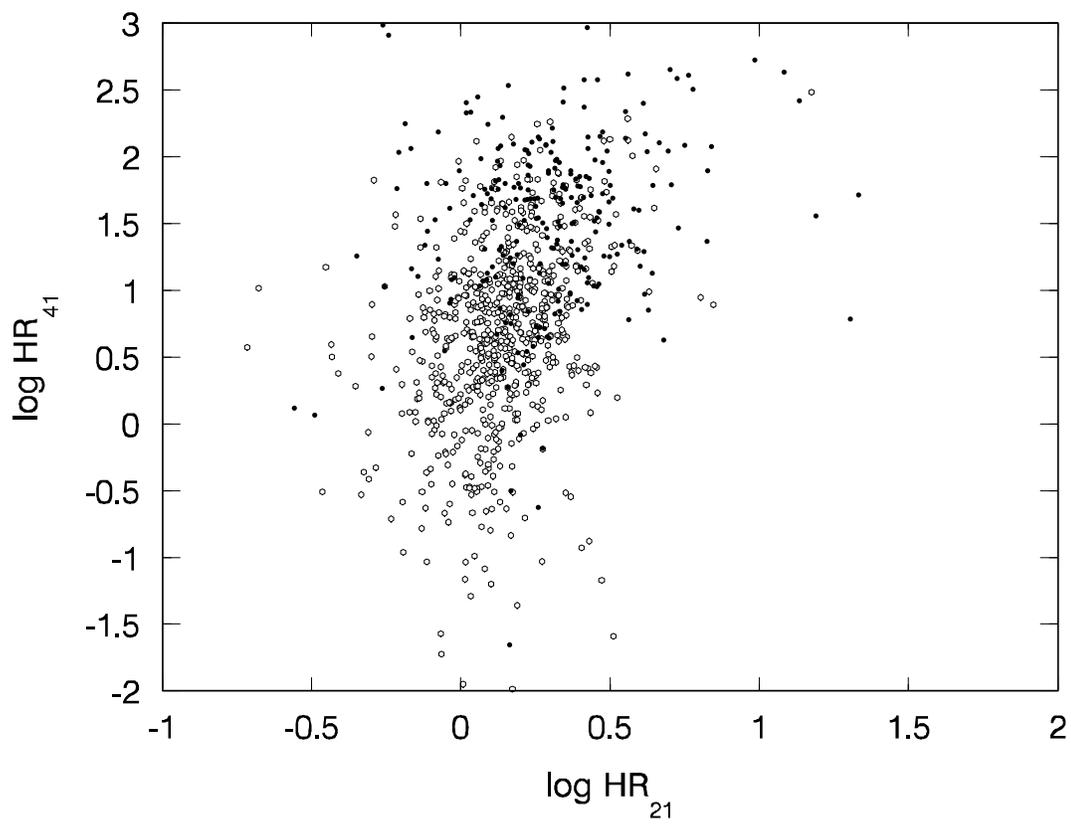
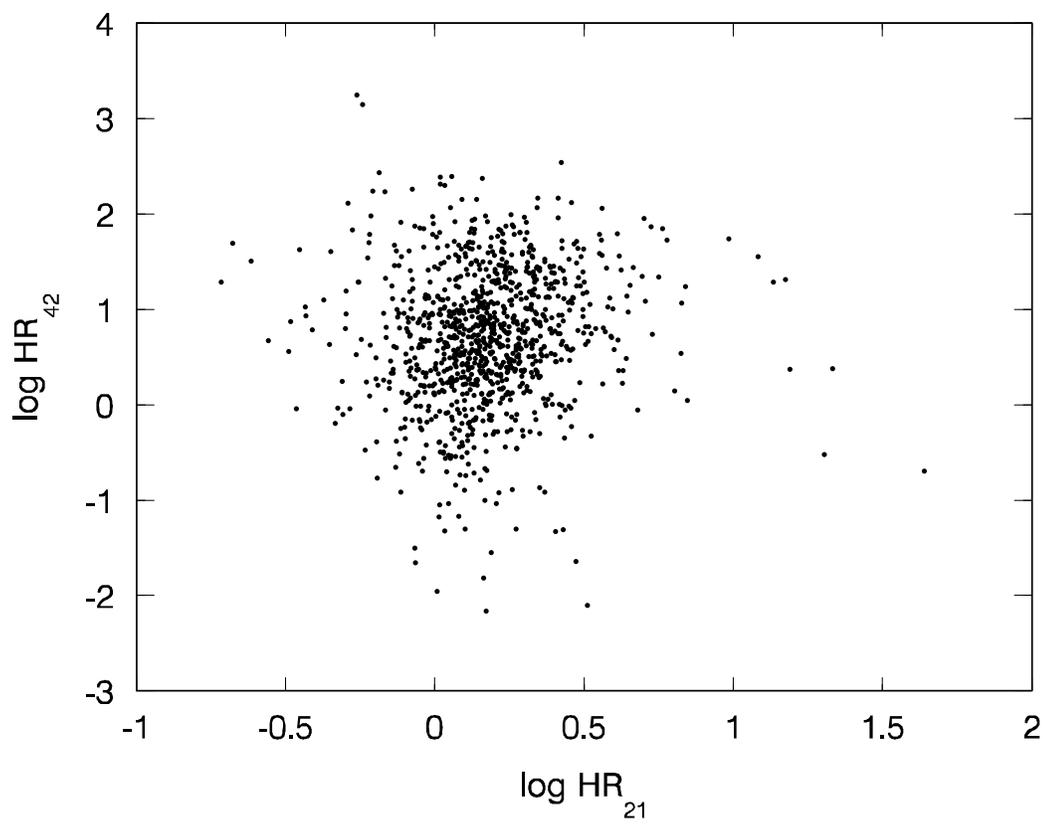
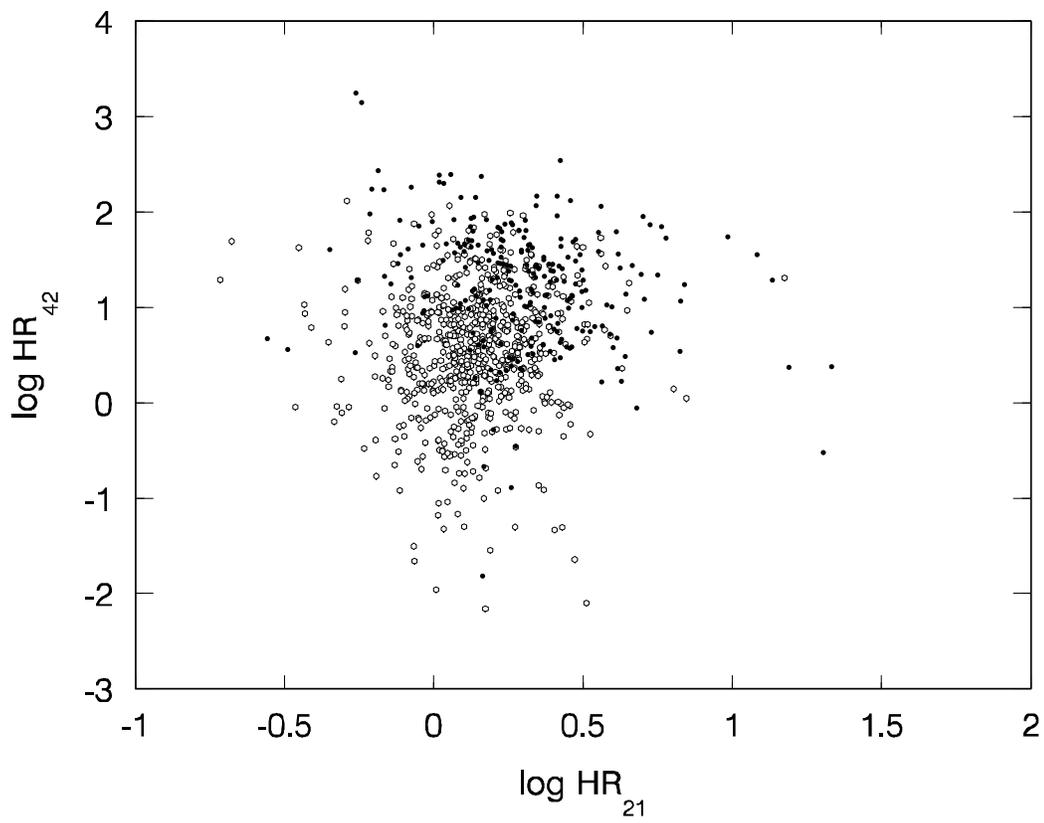
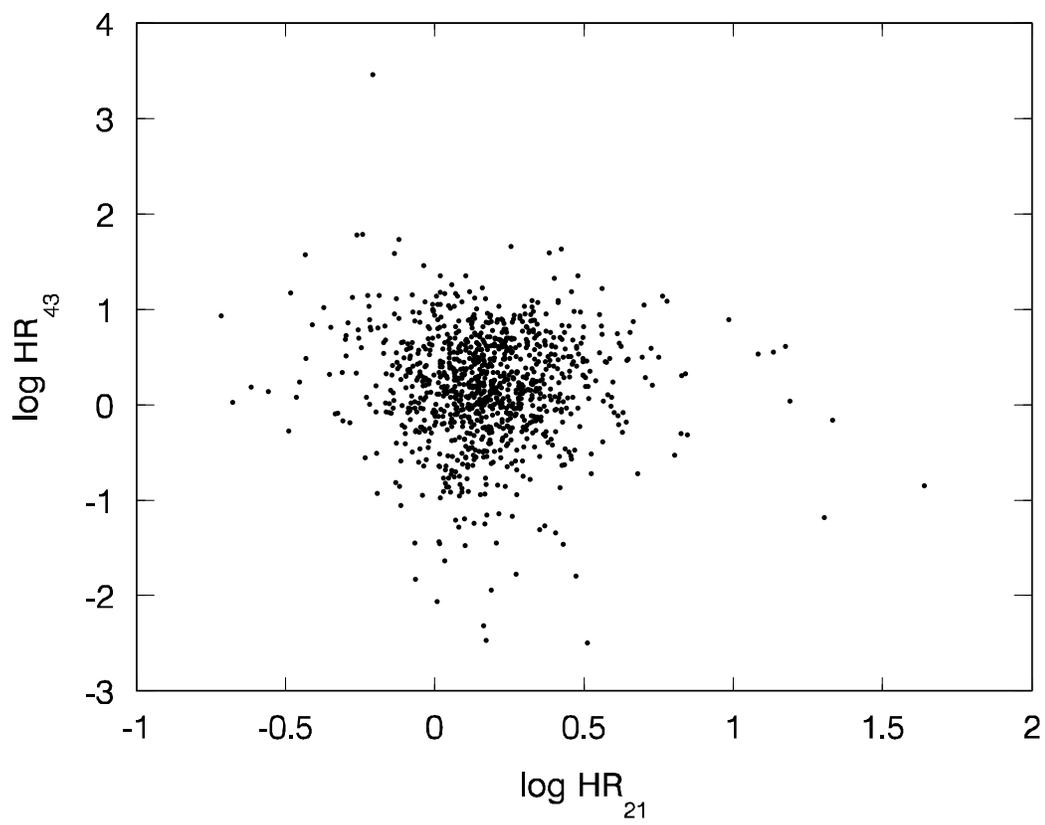
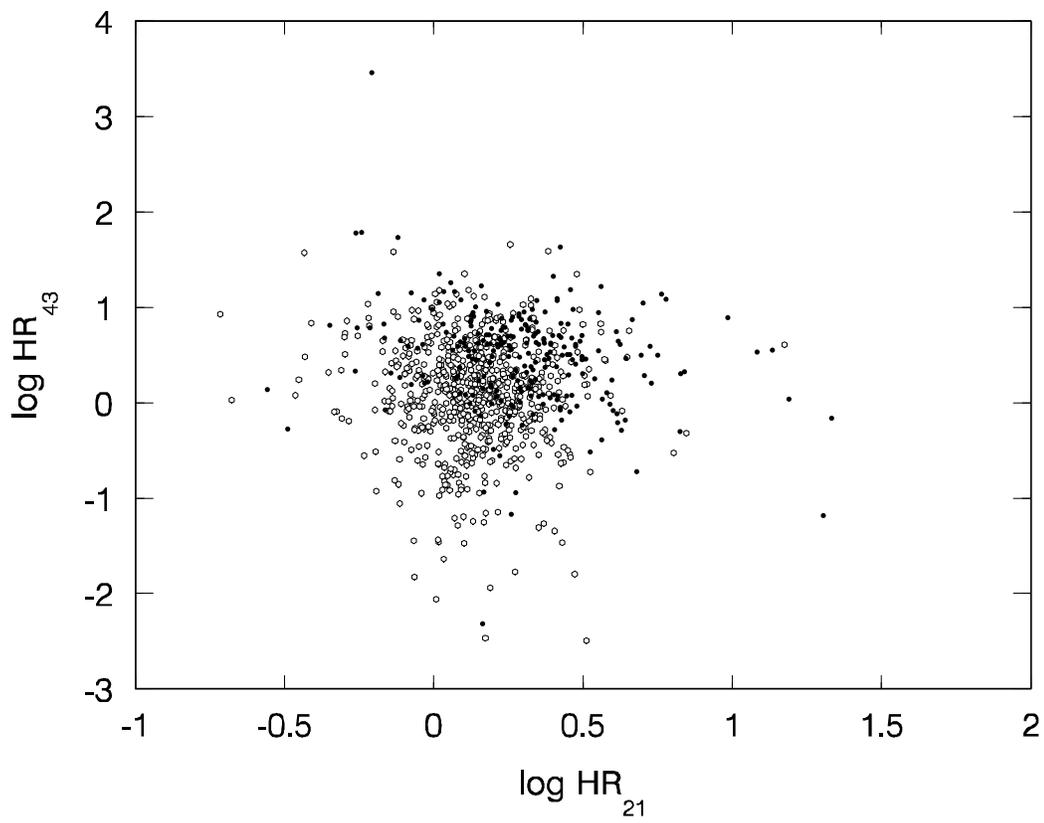
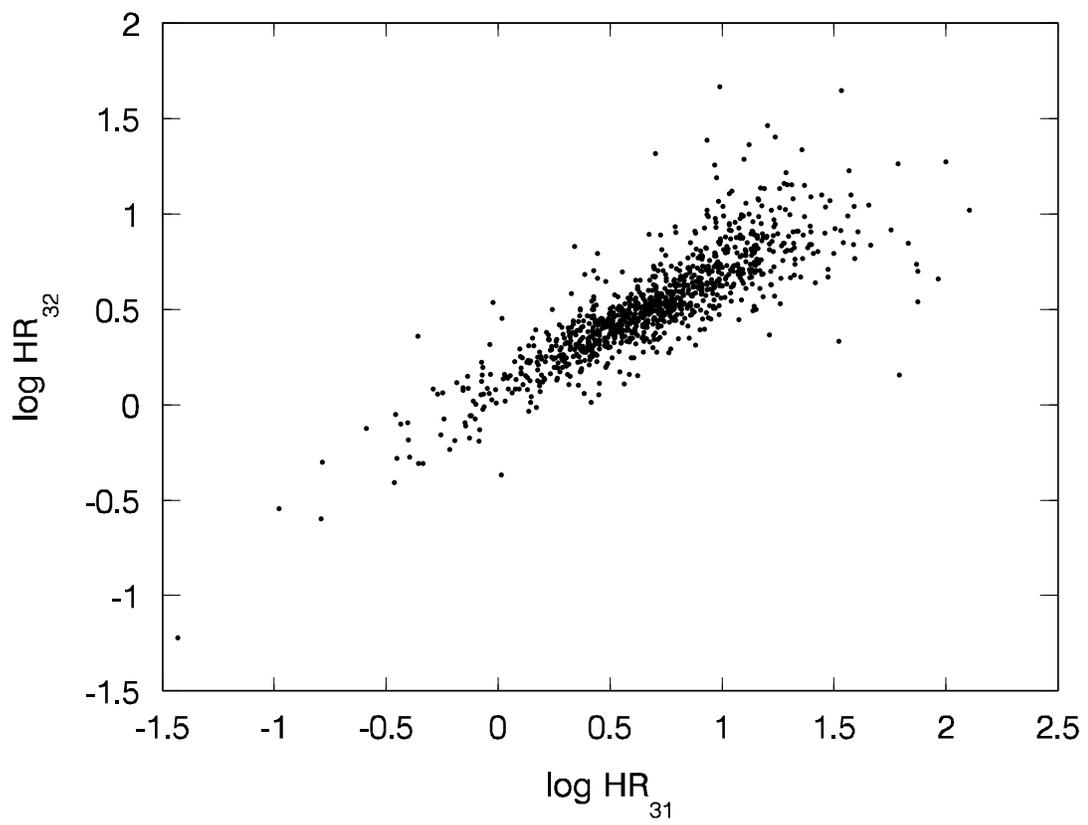
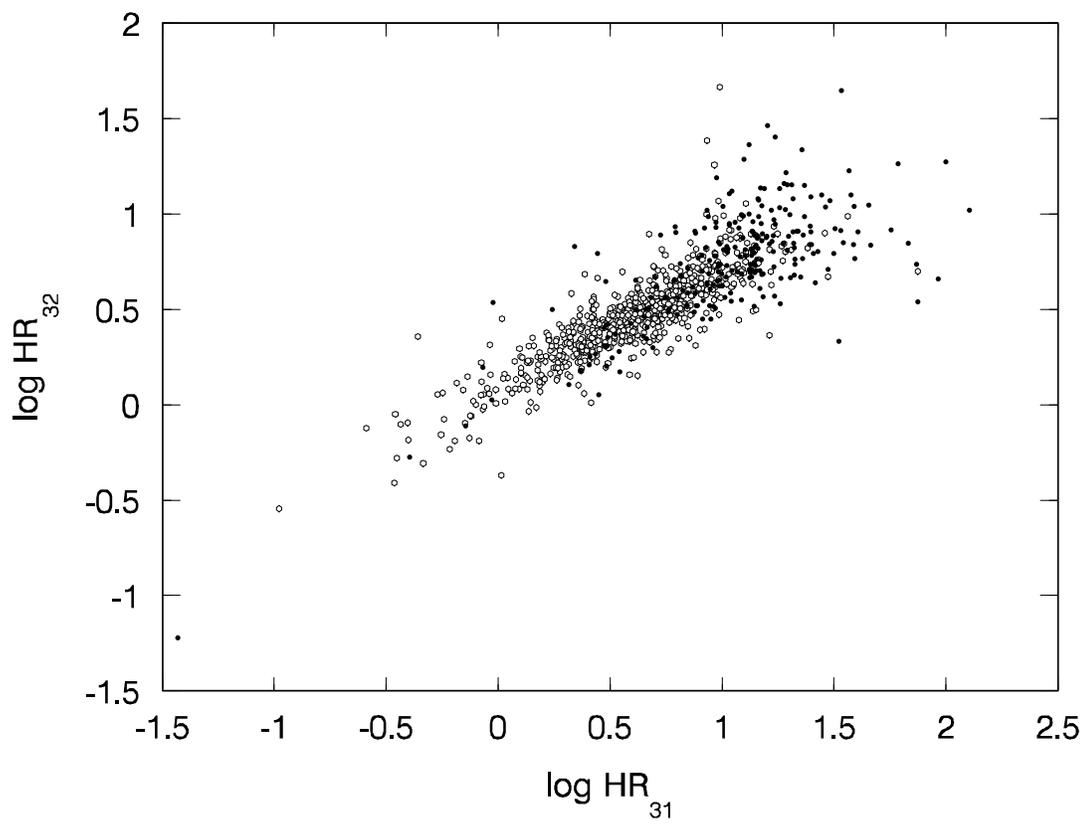
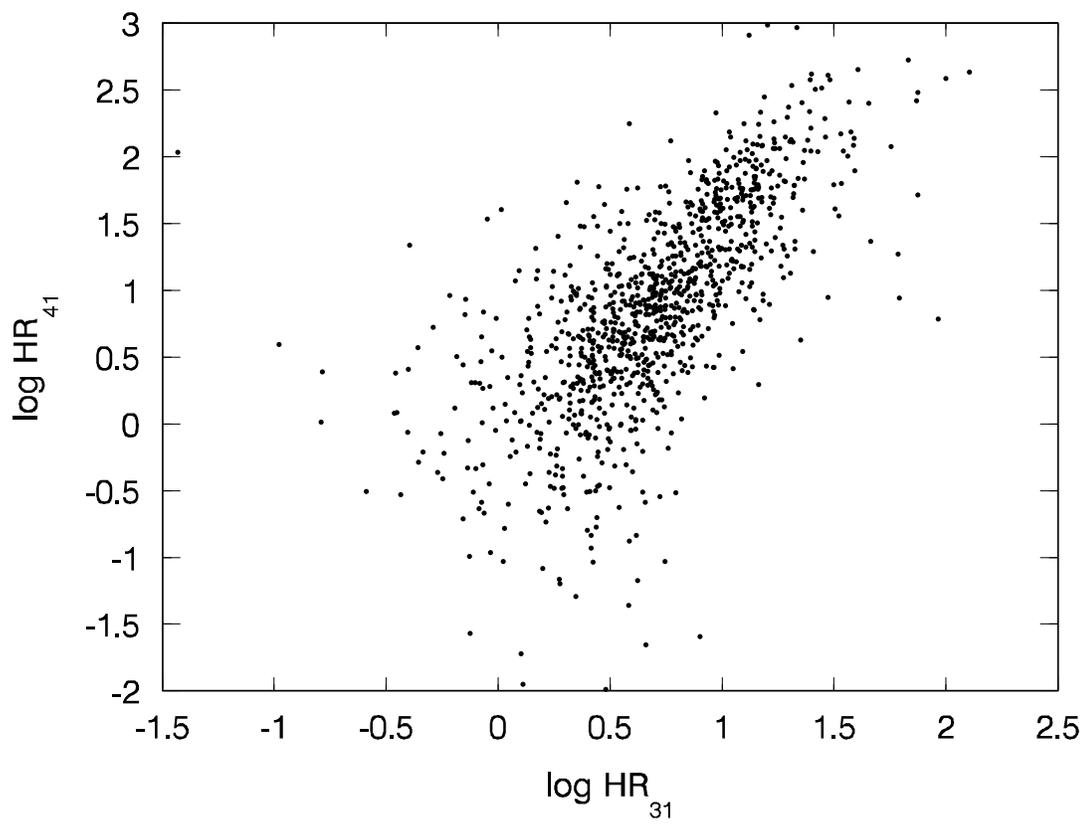
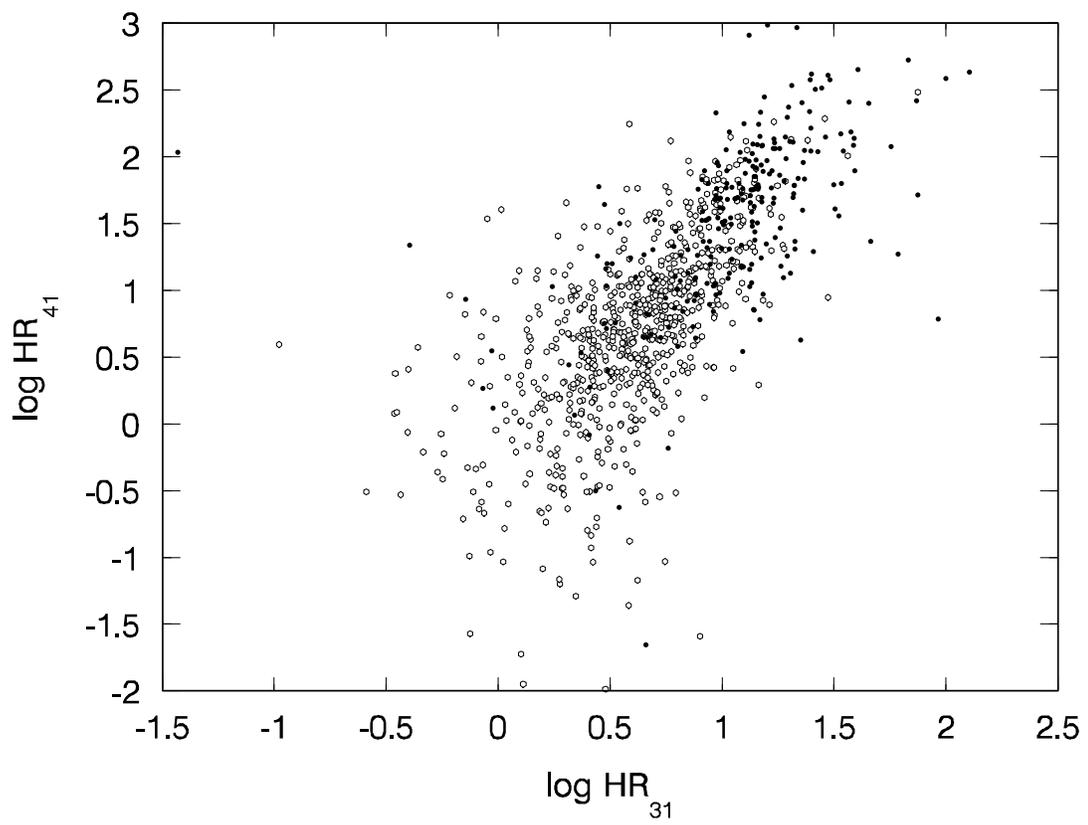
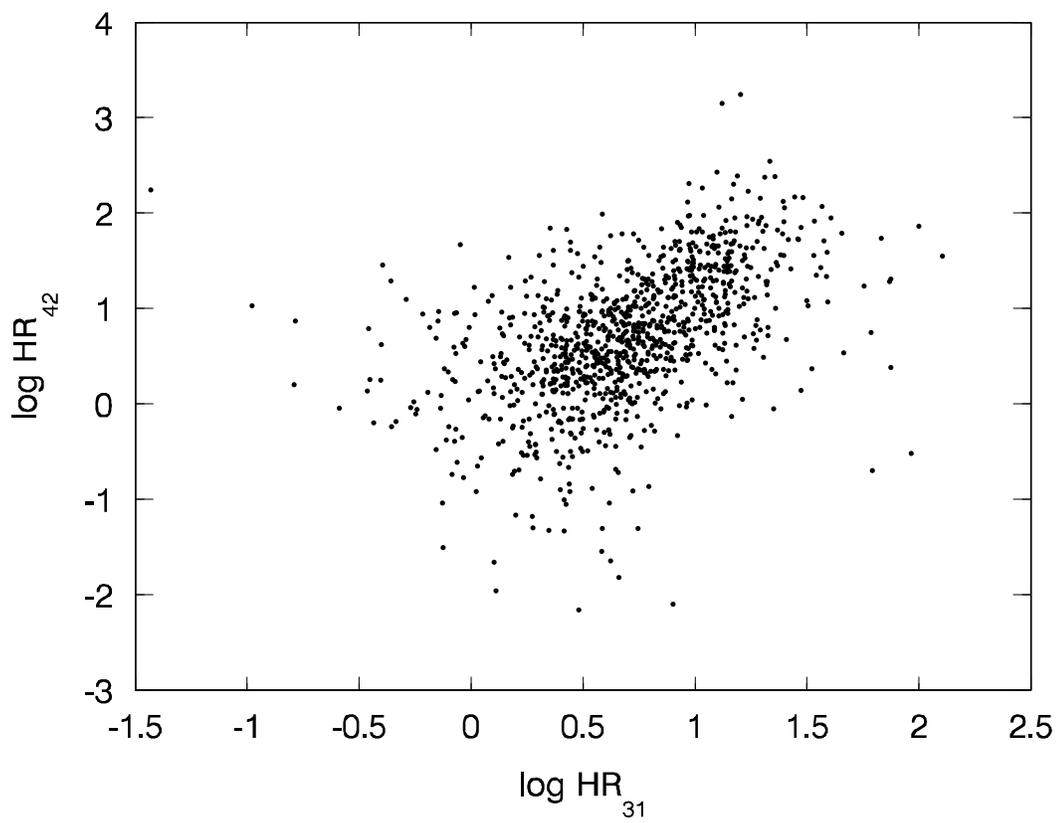
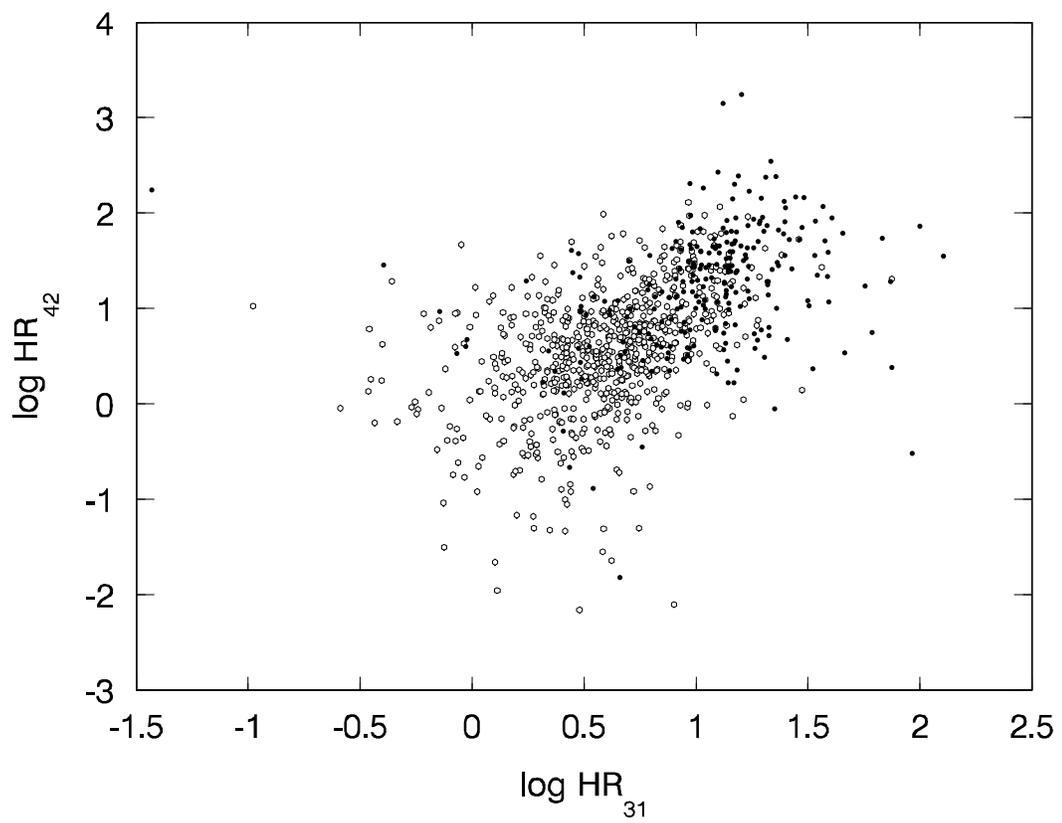
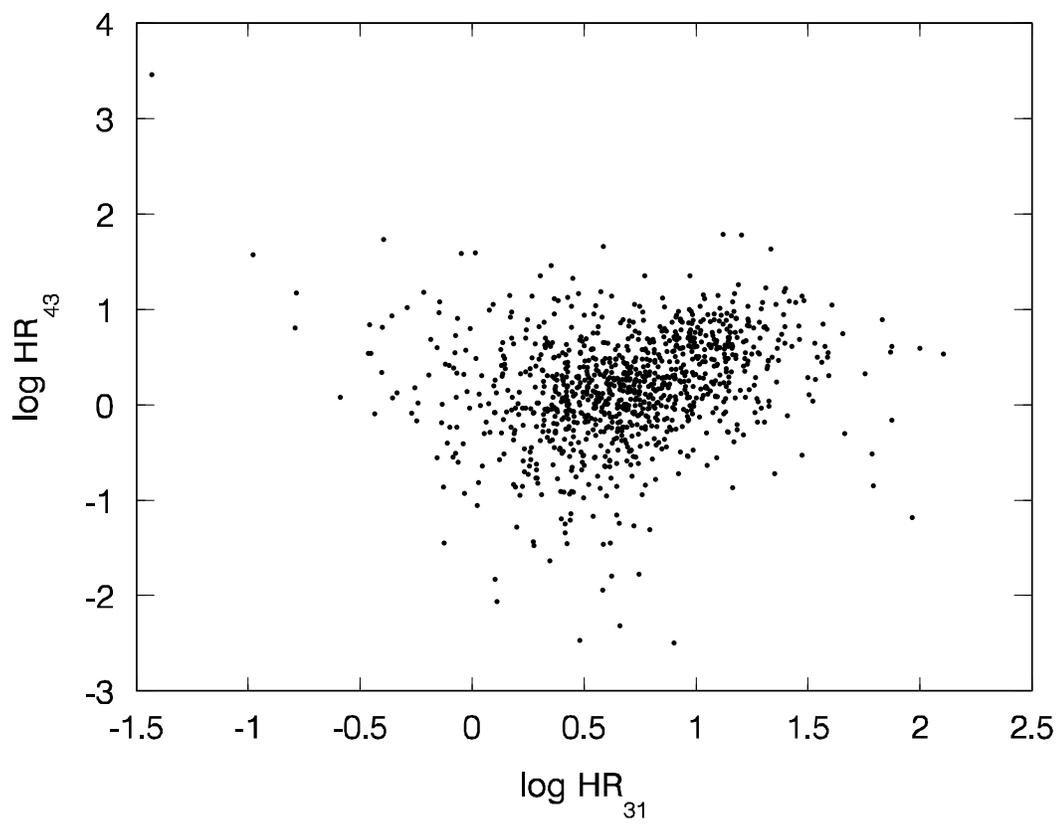
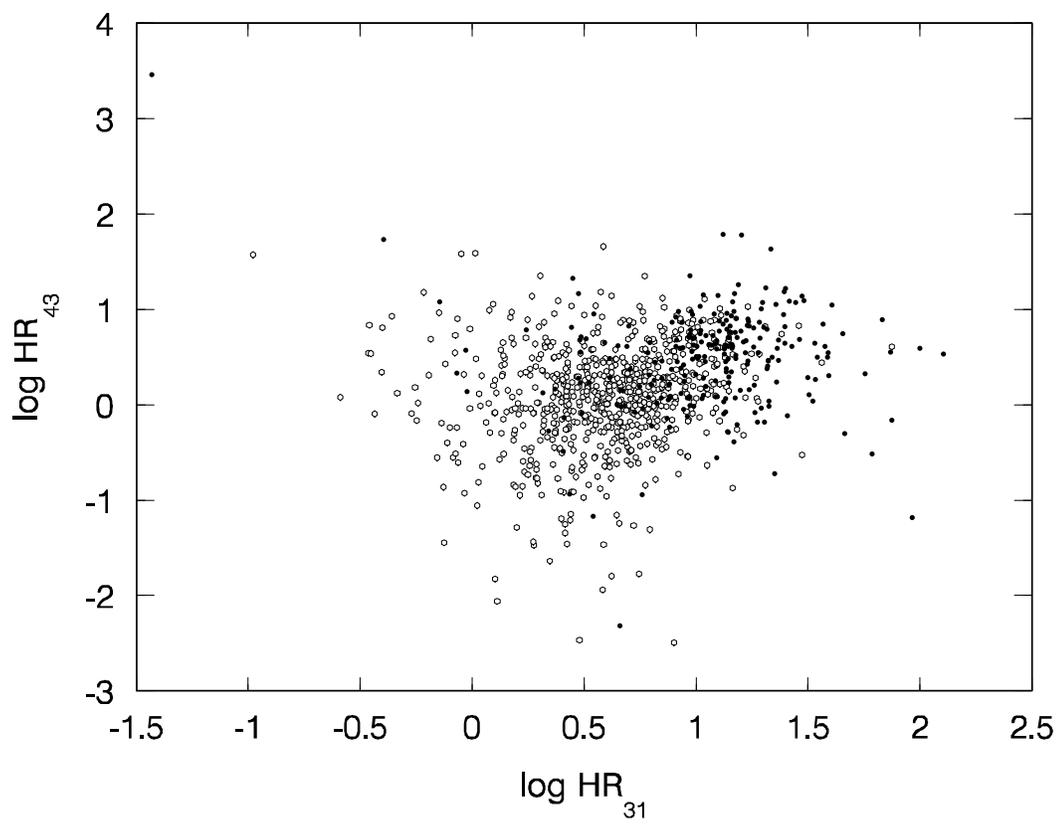
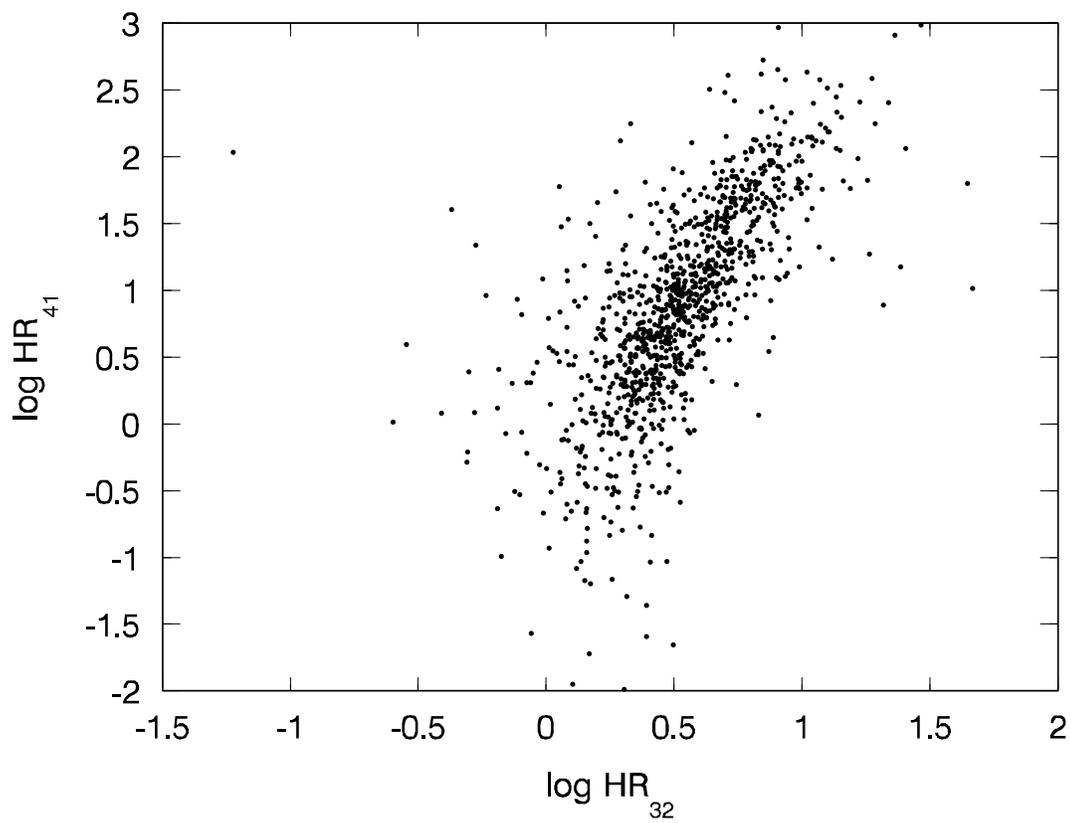
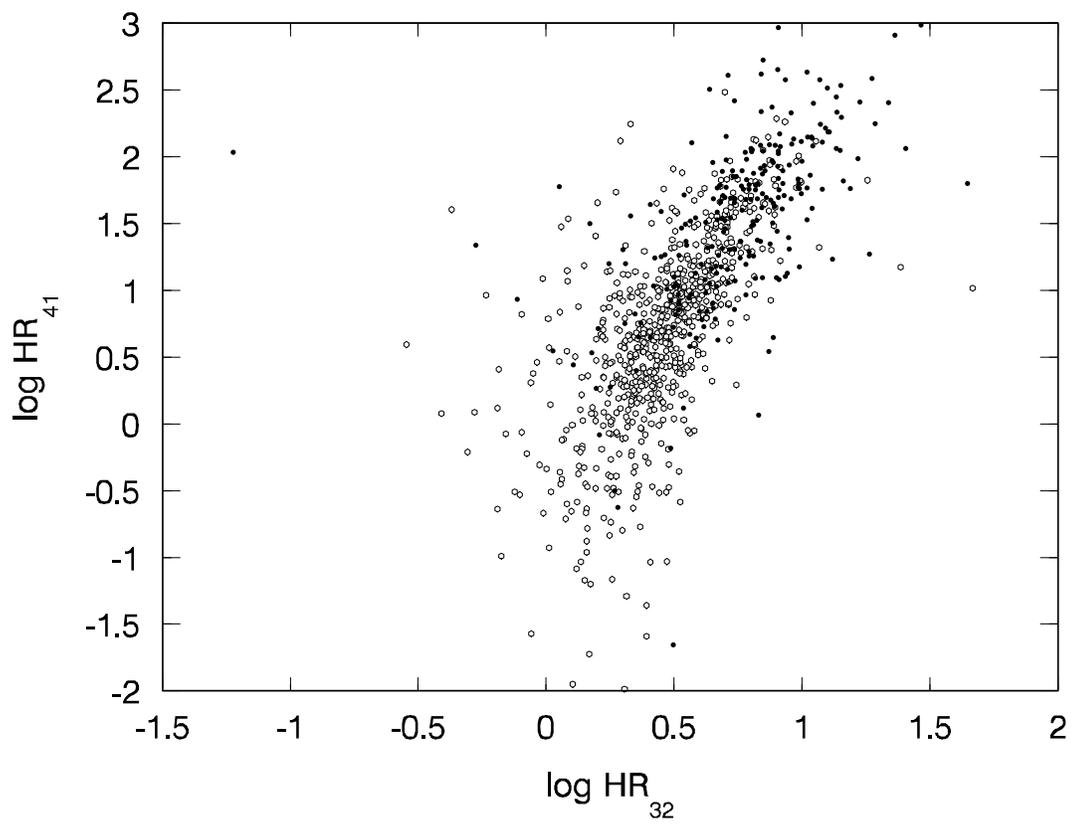
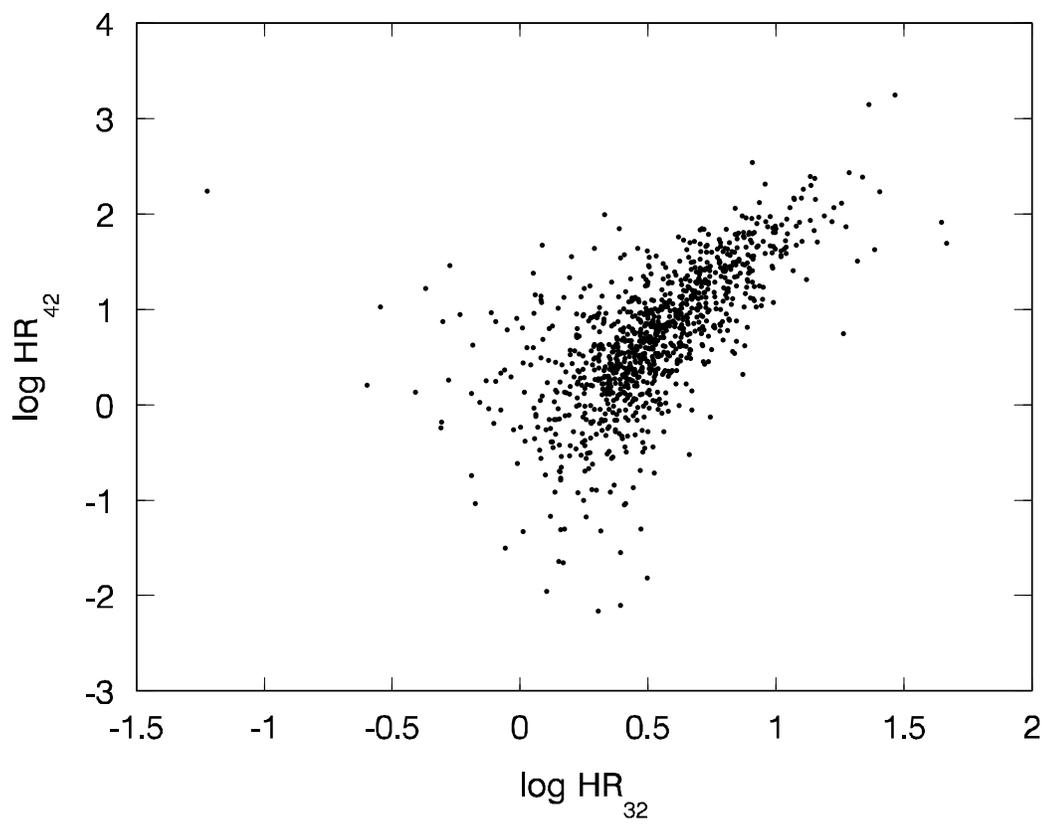
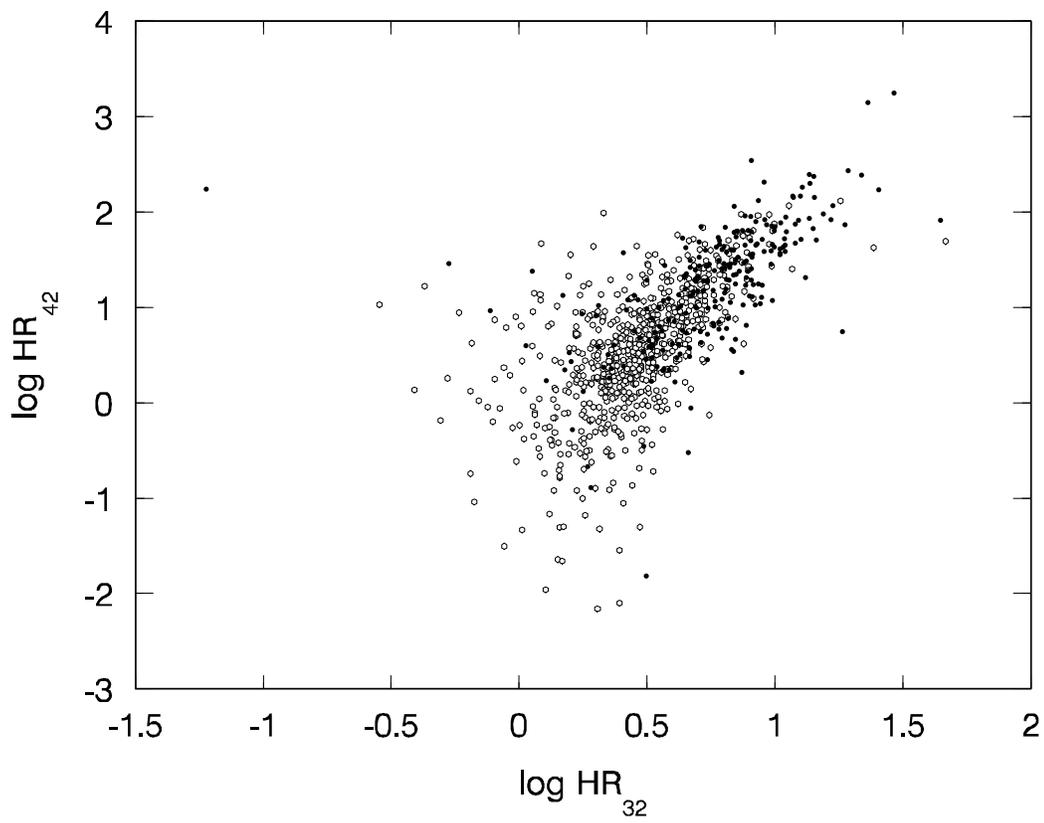
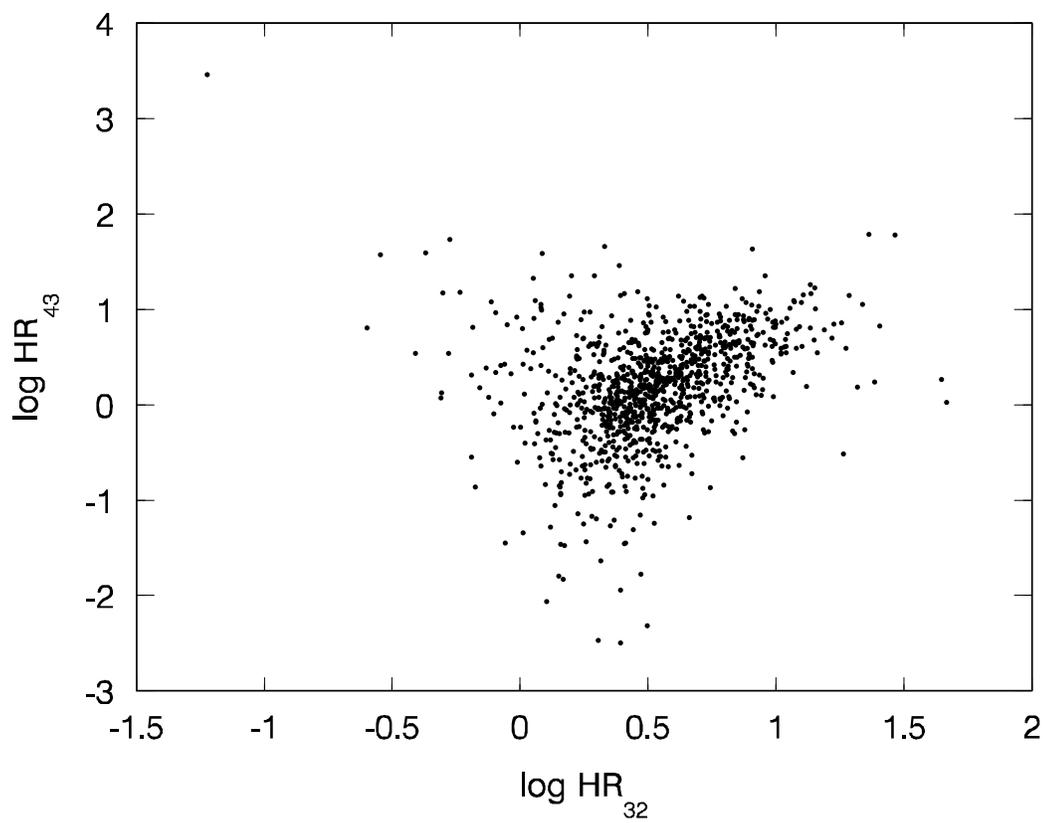
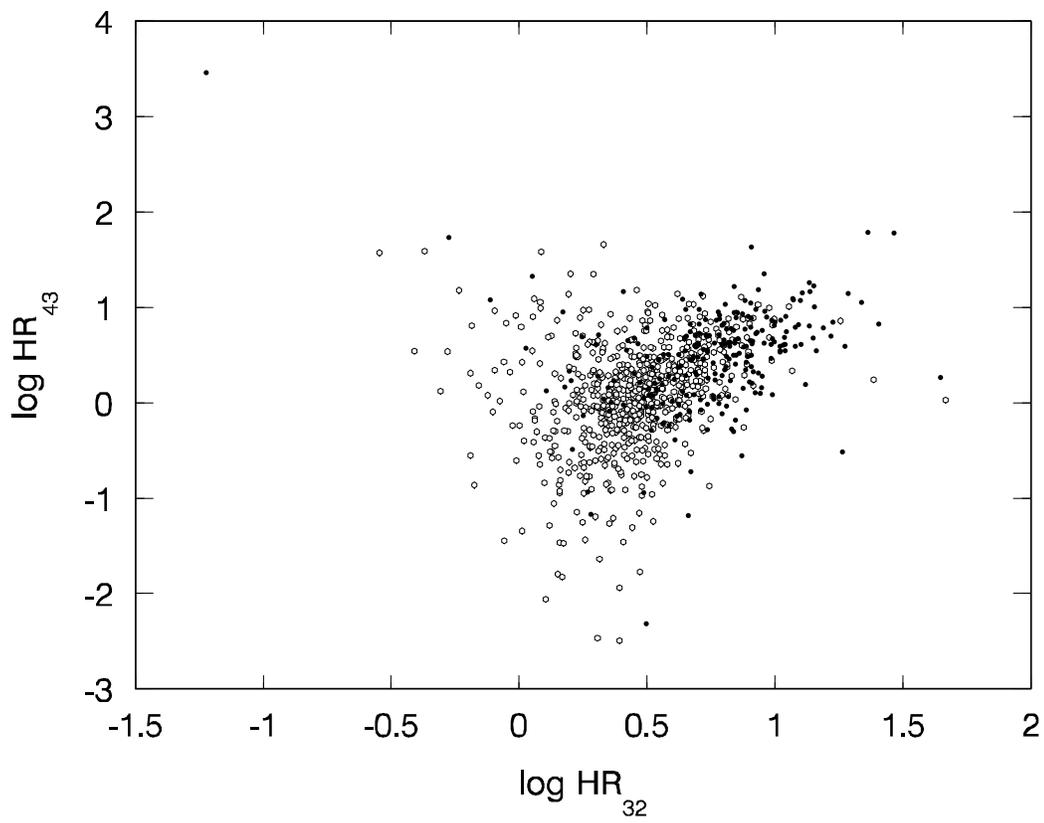
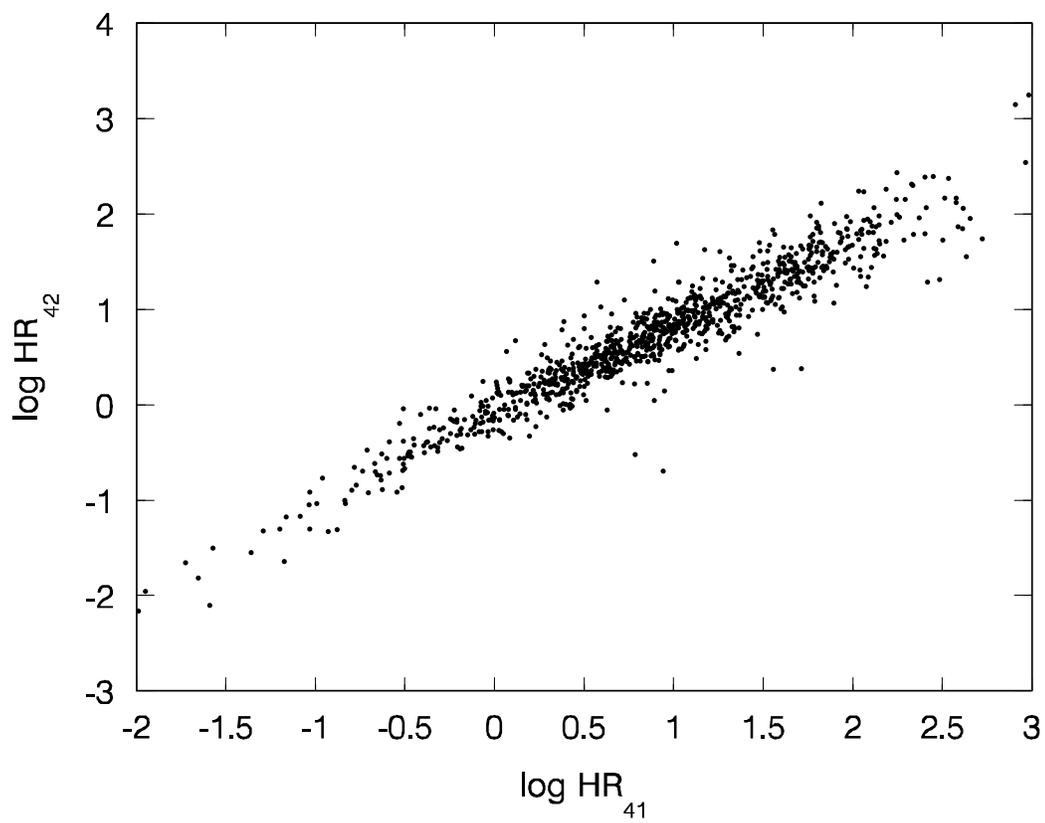
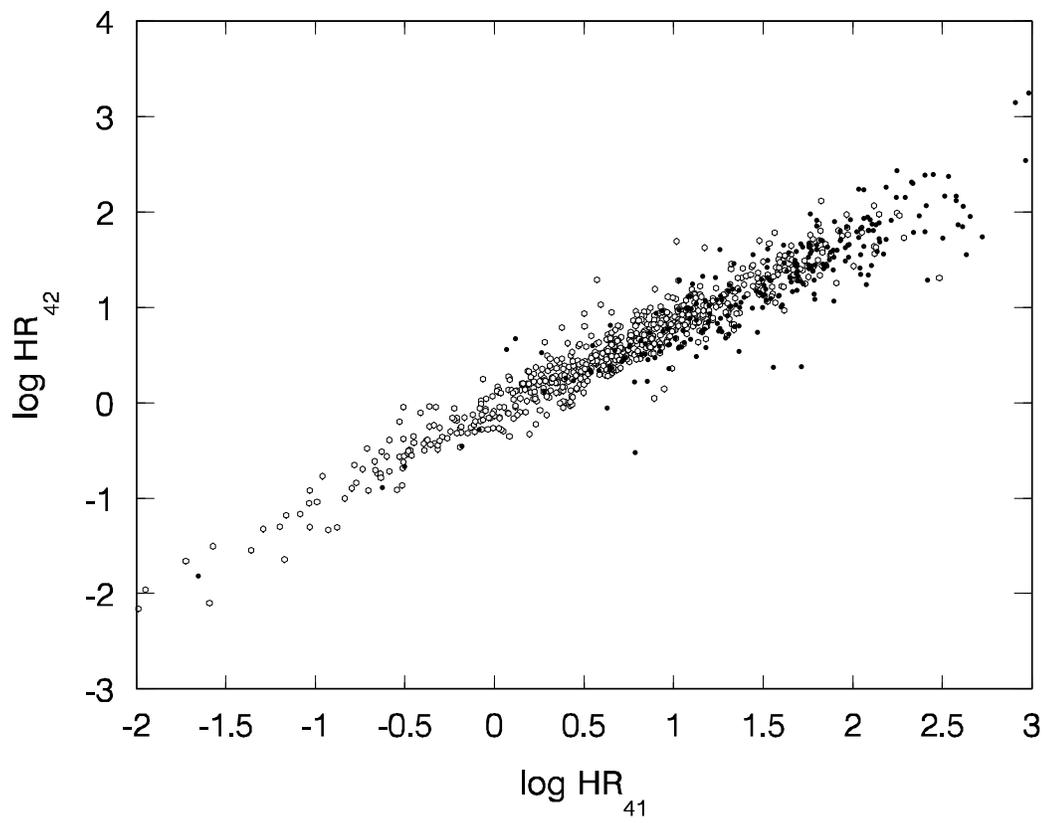
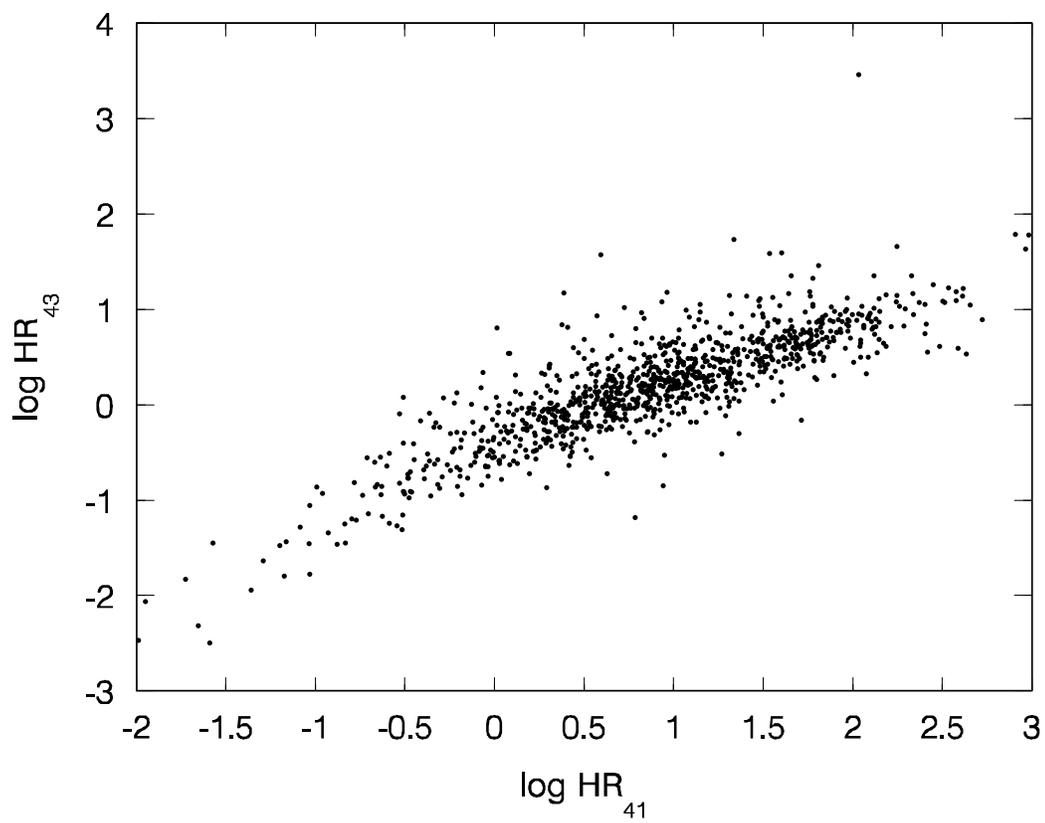
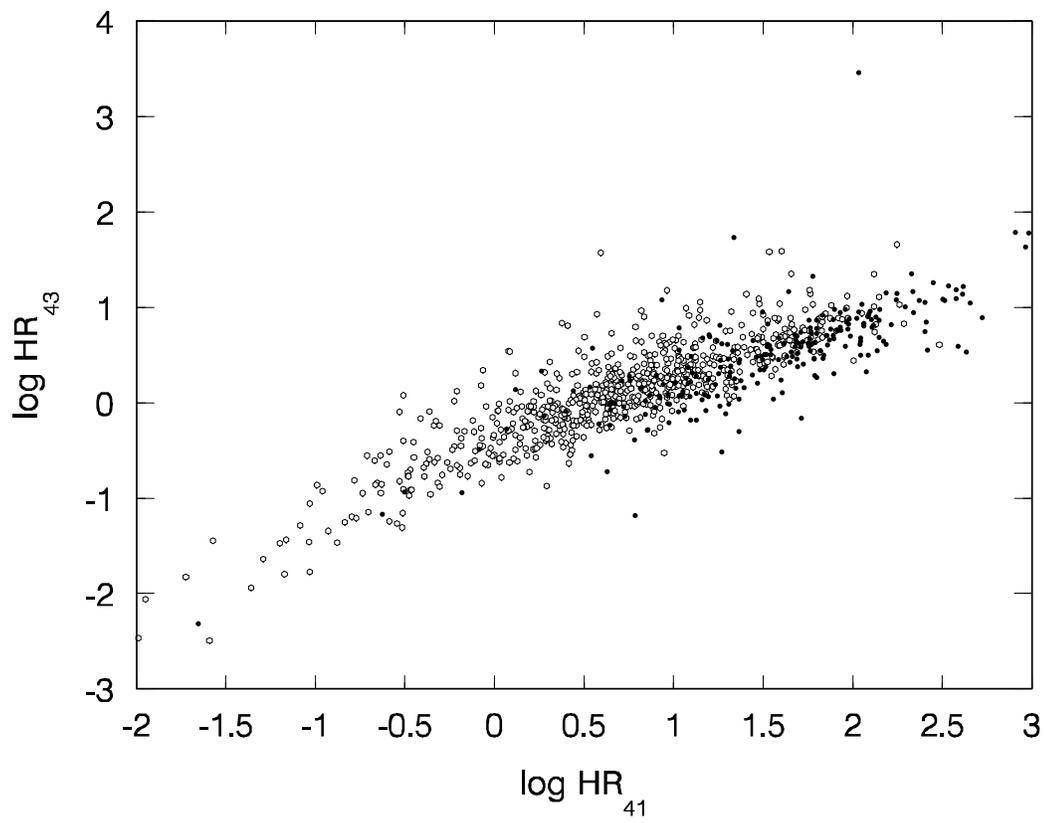
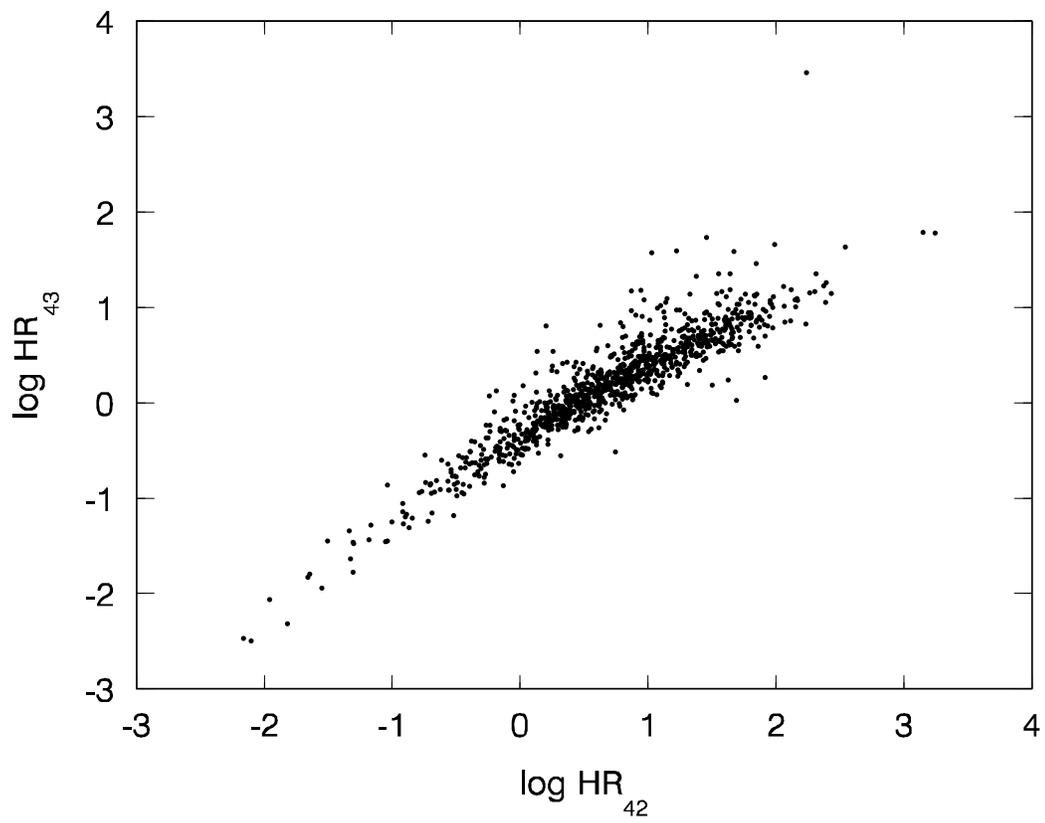
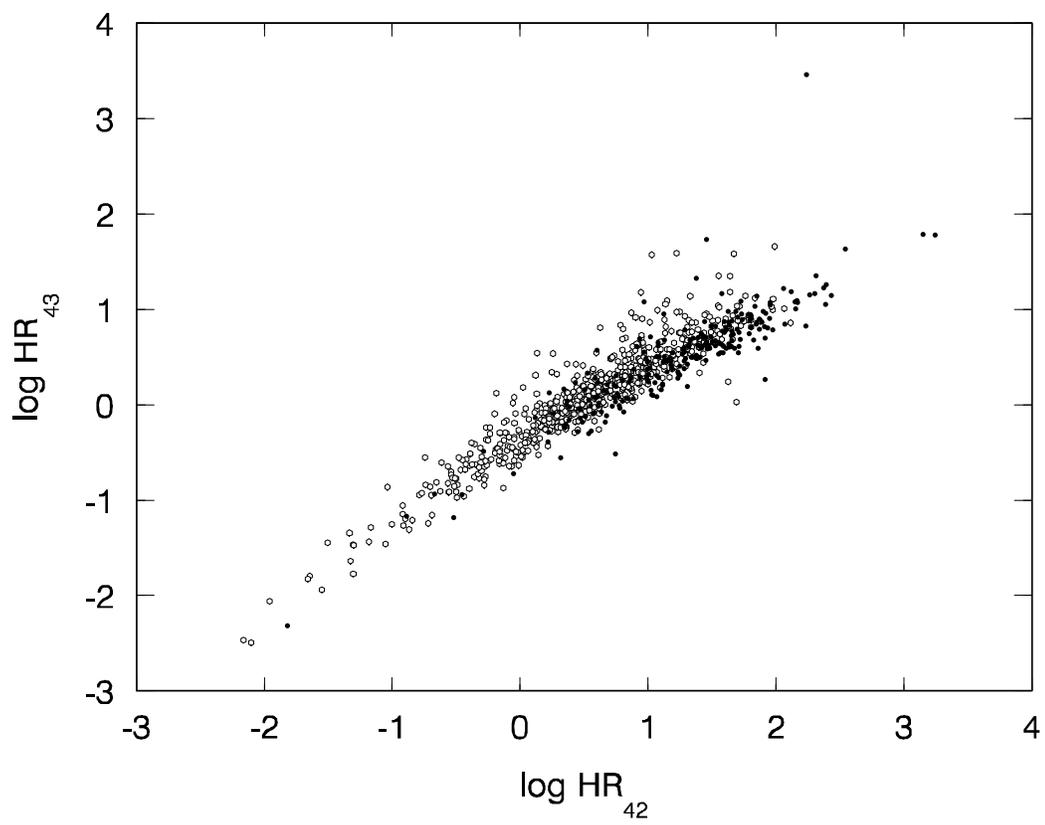
\end{document}